\documentstyle[preprint,aps,epsf]{revtex}
\begin{document}

\tightenlines

\title{The properties of $\bar{K}$ in the nuclear medium}

\author{A. Ramos}
\address{Departament d'Estructura i Constituents de la Mat\`eria,
Universitat de Barcelona, \\
Diagonal 647, 08028 Barcelona, Spain
}

\author{E. Oset}

\address{
Departamento de F\'{\i}sica Te\'orica and IFIC,
Centro Mixto Universidad de Valencia-CSIC, \\
46100 Burjassot (Valencia), Spain}

\date{\today}

\maketitle
\begin{abstract}

The self-energy of the $K^-$ meson in nuclear matter is
calculated in a
self-consistent microscopic approach, using a $\bar{K}N$
interaction obtained
from the lowest-order meson-baryon chiral Lagrangian. The
effective $\bar{K}N$
interaction in the medium is derived by solving the
coupled-channel
Bethe-Salpeter equation including Pauli blocking on the
nucleons, mean-field
binding potentials for the baryons and the self-energy of the
$\pi$ and $\bar{K}$
mesons.  The incorporation of the self-consistent ${\bar K}$
self-energy in the
description, in addition to the Pauli blocking effects, yields a
weaker
attractive in-medium ${\bar K}N$ interaction and a
$\Lambda(1405)$ which dissolves
faster with increasing matter density, as a result of the ${\bar
K}$ spectral
function being spread out over a wide range of energies. These
effects are
further magnified when the intermediate pions are dressed.


\end{abstract}
\vskip 0.5 cm



\section{Introduction}
\label{sec:intro}

The properties of the kaons and antikaons in the nuclear medium
have been the object of numerous investigations since the
possibility of the existence of a kaon
condensed phase in dense nuclear matter was pointed out
\cite{KN86}.
The enhancement of the $K^-$ yield in Ni+Ni collisions measured
recently by the KaoS
collaboration at
GSI \cite{kaos97} can be explained by assuming the $K^-$ meson to
feel a strong
attraction in the medium \cite{cassing97,Li97}. Kaonic atom data, 
a compilation of which is given in Ref. \cite{FGB94},
also favor an attractive $K^-$ nucleus interaction.

Most of the recent theoretical works start from chiral Lagrangians
that
reproduce the free space scattering properties which, in the case
of $K^- p$
scattering, are dominated by the presence of an isospin zero
resonance, the
$\Lambda(1405)$.
It is precisely the influence of this resonance what makes the
$K^-p$ interaction repulsive at threshold, while the lowest order
Born s-wave amplitude from the chiral Lagrangian (the
Tomozawa-Weinberg term) is attractive. Evidently, one cannot
expect chiral
perturbation theory to work close to a resonance. This is why all
works studying the free space $K^-N$ interaction from the chiral
Lagrangian
either introduce the $\Lambda(1405)$ as an elementary field
\cite{Lee9596}
or generate it dynamically through the Lippmann-Schwinger \cite{Kai9597} or
the Bethe-Salpeter \cite{oset98} equations.
These latter approaches allow for a
microscopic
incorporation
of the medium effects on the
the ${\bar K}N$ interaction \cite{Koch94,WKW96,Waas97,Lutz98}.
For instance, Pauli blocking on the intermediate nucleon
states, makes the
${\bar K}N$ interaction density dependent and this, in turn,
modifies the $K^-$
properties from those in free space. These medium modifications
were already included long time ago in the context of 
Brueckner-type many body theory \cite{alberg76} to obtain the 
kaon-nucleus optical potential for kaonic atoms. 
The medium properties of kaons and antikaons have also been
obtained from 
mean field theories, built within the framework of chiral
Lagrangians
\cite{Li97,Mao99}, based on the
relativistic Walecka-type model extended to incorporate
strangeness
in the form of hyperons or kaons \cite{Sch97}, or using explicitly
quark degrees of freedom \cite{Tsushi98}.

All the different approaches agree qualitatively in establishing
that, in the
medium, the $K^+$ feels a moderate repulsion and the $K^-$ a
strong attraction.
How large is this attraction is still 
somewhat controversial. Recent
phenomenological approaches \cite{FGB94} based on fits to kaonic
atom data find a
$K^-$-nucleus potential of the order of $-200 \pm 20$ MeV in the
nuclear center.
However, no calculation that starts from the bare $K^- N$
interaction predicts
such an attraction, the values ranging from $-140$ to $-75$ MeV.
Hopefully,
heavy-ion reactions, that are sensitive to higher density
regions, will help in
elucidating these discrepancies. At the same time, it is
necessary to develop
theories that treat the intricacies associated to the mutual
interaction between
all the hadrons in the medium as accurately as possible.

In the present work we perform a microscopic study of the the
$K^-$ properties in
nuclear matter by incorporating the medium modifications on the
${\bar K}N$
amplitude using the model of Ref. \cite{oset98}, which 
was shown to reproduce the $K^-p$ scattering
observables
very satisfactorily. As mentioned above, one source
of density
dependence is the Pauli blocking on the nucleon states. This makes
the
${\bar K}N$ interaction attractive and, on the other hand,
shifts the resonance
to higher energies \cite{Koch94,Waas97}.
However, it was shown recently \cite{Lutz98} that a
self-consistent calculation
of the  $K^-$ self-energy leaves the position of the resonance
unchanged, due to
a compensation of the repulsive $\Lambda(1405)$ shift with the
attraction
felt by the $K^-$ meson. The importance of these medium effects
makes it
interesting to investigate other medium modifications of the
particles participating in building up the ${\bar K}N$
interaction. This is the aim of the present work. To this end, we
include, in addition, the dressing
of the pions in the $\pi \Lambda$, $\pi \Sigma$ intermediate
states, which couple
strongly to the ${\bar K} N$ state. This is done through a pion
self-energy that
contains the effect of one- and two-nucleon absorption,
conveniently
modified to include the effect of nuclear short-range correlations.
The
medium effects on
the nucleons and hyperons is considered via density-dependent
mean-field binding
potentials.

After reviewing the free-space formalism in Sect. \ref{sec:free},
we describe,
in Sect.  \ref{sec:medium}, the appropiate
modifications needed to incorporate the medium effects on the
$\bar{K}N$ amplitude. The details of the kaon and pion
self-energies are given in Sect. \ref{sec:self}. Our results are
discussed in
Sect. \ref{sec:results}, where various approximations are
compared. Finally,
Sect. \ref{sec:conclusions} summarizes our main results and
conclusions.

\section{$\bar{K}N$ amplitudes in free space}
\label{sec:free}

In this section we review the formalism used in
Ref. \cite{oset98}
for describing $\bar{K}N$ scattering in free space.
The effective chiral Lagrangian formalism has been
very successful in
explaining the properties of meson-meson interaction at low
energies \cite{Ga85,Pi95},
as well as those of the meson-baryon systems
\cite{Eck95,Be95} when the interaction is weak, as in the case of
the s-wave $\pi N$ and $K^+ N$ interaction.
However, the $\bar{K} N$ system
couples strongly
to many other channels and generates a resonance below threshold,
the $\Lambda (1405)$. In this case the standard chiral
scheme,
an expansion in powers of the typical momenta involved in the
process, fails
to be an appropriate approach.

In Ref. \cite{oset98} a
non perturbative scheme, consisting of
solving a set of coupled-channel Bethe-Salpeter equations
using the lowest order chiral Lagrangian in S-wave, was shown
to reproduce the low energy $K^- p$ scattering data very
satisfactorily with only one-parameter, the cut-off used to
renormalize the loop integrals.
A reinterpretation of these equations to the light of the inverse
amplitude method can be seen in Ref. \cite{OOP98}, where it is
shown that the effect of higher order chiral Lagrangians can be
reabsorbed with the choice of a suitable cut-off in some cases.
The model of Ref. \cite{oset98} follows closely that of
Ref.~\cite{Kai9597},
where the success of the non-perturbative approach using the
chiral
Lagrangians was first shown, but includes the complete set of
$0^-$ meson and $1/2^+$
baryon octet states in the basis space of coupled channels. The use of this
complete set, apart from providing the right
SU(3)
symmetry in the limit of equal baryon masses and equal meson
masses,
was found to be essential for reproducing the experimental
branching ratios at threshold
with only the lowest order Lagrangian.

The lowest order chiral
Lagrangian,
coupling the octet of pseudoscalar mesons to the octet of $1/2^+$
baryons, is
\begin{eqnarray}
L_1^{(B)} = &&\langle \bar{B} i \gamma^{\mu} \nabla_{\mu} B
\rangle -
M_B \langle \bar{B} B \rangle \nonumber \\
& + &\frac{1}{2} D \langle \bar{B} \gamma^{\mu} \gamma_5 \left\{
u_{\mu},
B \right\} \rangle
+ \frac{1}{2} F \langle \bar{B} \gamma^{\mu} \gamma_5 [u_{\mu},
B]
\rangle \ ,
\end{eqnarray}
where the symbol $\langle\, \rangle$ denotes the trace of SU(3)
matrices
and
\begin{equation}
\begin{array}{l}
\nabla_{\mu} B = \partial_{\mu} B + [\Gamma_{\mu}, B] \\
\Gamma_{\mu} = \frac{1}{2} (u^\dagger \partial_{\mu} u + u
\partial_{\mu} u^\dagger) \\
U = u^2 = {\rm exp} (i \sqrt{2} \Phi / f) \\
u_{\mu} = i u ^\dagger \partial_{\mu} U u^\dagger  \ .
\end{array}
\end{equation}
The SU(3) matrices for the mesons and the baryons are the
following
\begin{equation}
\Phi =
\left(
\begin{array}{ccc}
\frac{1}{\sqrt{2}} \pi^0 + \frac{1}{\sqrt{6}} \eta & \pi^+ & K^+
\\
\pi^- & - \frac{1}{\sqrt{2}} \pi^0 + \frac{1}{\sqrt{6}} \eta &
K^0 \\
K^- & \bar{K}^0 & - \frac{2}{\sqrt{6}} \eta
\end{array}
\right) \ ,
\end{equation}

\begin{equation}
B =
\left(
\begin{array}{ccc}
\frac{1}{\sqrt{2}} \Sigma^0 + \frac{1}{\sqrt{6}} \Lambda &
\Sigma^+ & p \\
\Sigma^- & - \frac{1}{\sqrt{2}} \Sigma^0 + \frac{1}{\sqrt{6}}
\Lambda & n \\
\Xi^- & \Xi^0 & - \frac{2}{\sqrt{6}} \Lambda
\end{array}
\right) \ .
\end{equation}
At lowest order in momentum the
interaction
Lagrangian reduces to
\begin{equation}
L_1^{(B)} = \langle \bar{B} i \gamma^{\mu} \frac{1}{4 f^2}
[(\Phi \partial_{\mu} \Phi - \partial_{\mu} \Phi \Phi) B
- B (\Phi \partial_{\mu} \Phi - \partial_{\mu} \Phi \Phi)]
\rangle     \ .
\end{equation}
The coupled channel formalism requires to evaluate the transition
amplitudes between the different channels that can be built
from
the meson and baryon octets. For $K^- p$ scattering there are 10
channels, namely $K^-p$, $\bar{K}^0 n$, $\pi^0
\Lambda$, $\pi^0 \Sigma^0$,
$\pi^+ \Sigma^-$, $\pi^- \Sigma^+$, $\eta \Lambda$, $\eta
\Sigma^0$,
$K^+ \Xi^-$ and $K^0 \Xi^0$. In the case of $K^- n$ scattering
the coupled channels are: $K^-n$, $\pi^0\Sigma^-$,
 $\pi^- \Sigma^0$, $\pi^- \Lambda$, $\eta
\Sigma^-$ and
$K^0 \Xi^-$.
These amplitudes have the form
\begin{equation}
V_{i j} = - C_{i j} \frac{1}{4 f^2} \bar{u} (p_i) \gamma^{\mu} u
(p_j)
(k_{i \mu} + k_{j \mu}) \ ,
\label{eq:v-matrix}
\end{equation}
where $p_j, p_i (k_j, k_i)$ are the initial, final momenta of the
baryons (mesons).
The explicit values of the coefficients $C_{ij}$ can be
found in Ref. \cite{oset98}. At low energies we can neglect the
spatial components and Eq.~(\ref{eq:v-matrix}) simplifies to
\begin{equation}
V_{i j} = - C_{i j} \frac{1}{4 f^2} (k_j^0 + k_i^0) \ .
\end{equation}

The coupled-channel Bethe-Salpeter
equations in the center of mass frame read
\begin{equation}
T_{i j} = V_{i j} + \overline{V_{i l} \; G_l \; T_{l j}} \ ,
\end{equation}
\noindent
where the indices $i,l,j$ run over all possible channels and
\begin{equation}
\overline{V_{i l} \; G_l \; T_{l j}} = i \int \frac{d^4 q}{(2
\pi)^4}
\, \frac{M_l}{E_l (-\vec{q}\,)}
\, \frac{V_{i l} (k_i, q) \, T_{l j} (q, k_j)}{k^0 + p^0 - q^0 -
E_l
(-\vec{q}\,)
+ i \epsilon} \, \frac{1}{q^2 - m^2_l + i \epsilon} \ ,
\label{eq:loop}
\end{equation}
with
$M_l, E_l$ and $m_l$ being, respectively, the
baryon mass, baryon energy and meson mass in the intermediate
state.

Although Eq.~(\ref{eq:loop}) requires the half-off-shell
amplitudes, it was shown in Ref. \cite{oset98} that the
off-shell part goes into
renormalization of coupling constants.
Therefore, one can factorize the remainig on-shell components of
$V$ and $T$ outside the integral of
Eq. (\ref{eq:loop}) reducing the integral equation
to a set of algebraic equations. The loop integral reads
\begin{eqnarray}
G_{l}(\sqrt{s}) &=& i \, \int \frac{d^4 q}{(2 \pi)^4} \,
\frac{M_l}{E_l
(-\vec{q}\,)} \,
\frac{1}{\sqrt{s} - q^0 - E_l (-\vec{q}\,) + i \epsilon} \,
\frac{1}{q^2 - m^2_l + i \epsilon} \nonumber \\
&=& \int_{\mid {\vec q} \mid < q_{\rm max}} \, \frac{d^3 q}{(2
\pi)^3} \,
\frac{1}{2 \omega_l (\vec q\,)}
\,
\frac{M_l}{E_l (-\vec{q}\,)} \,
\frac{1}{\sqrt{s}- \omega_l (\vec{q}\,) - E_l (-\vec{q}\,) + i
\epsilon}
\label{eq:gprop}
\end{eqnarray}
with $\sqrt{s} = p^0 + k^0$.

The model discussed here depends on the loop regularization
cut-off, whose value $q_{\rm max}=630$ MeV was
chosen to reproduce the $K^- p$ scattering branching ratios
at threshold \cite{oset98}. The weak decay constant is $f=1.15
f_\pi$, a value
lying in between the pion and kaon ones that was chosen to optimize
the position of the $\Lambda(1405)$ resonance.
In Table \ref{tab:1} we summarize the predictions of the model for
several scattering observables at threshold. The scattering cross
sections, which are not used in the fit, were shown to be in good
agreement with the low energy data \cite{oset98} and the model
can safely be used up to 500 MeV/c $K^-$ lab momentum
\cite{nacher98} up to some punctual discrepancy in the small 
$K^- p \to \bar{K}^0 n$ cross section around 400 MeV/c, where the
D-wave $\Lambda(1520)$ resonance, not accounted for in the theory,
shows up in the data.

\section{$\bar{K}$ in the nuclear medium}
\label{sec:medium}

The dynamics of the ${\bar K} N$ interaction at low energies is
dominated by the $\Lambda(1405)$ resonance, which in the approaches
of Refs. \cite{Kai9597,oset98} is interpreted as an isospin $I=0$
quasi-bound
$K^- p$ state. The $\Lambda(1405)$ resonance is slightly below the
$K^-p$ threshold and leads to a repulsive $K^-p$ amplitude.
Although the isospin $I=1$ $K^-n$ channel is attractive
\cite{Adm81}, the $K^-p,K^-n$ averaged amplitude, $\bar{T}$, is
still repulsive and, according to the low energy theorem, the $K^-$
self-energy ($\bar{T}\rho$) should also be repulsive.
However, kaonic atom data suggest that, even at a small fraction of
the normal nuclear matter density $\rho_0$, the $K^-$ feels a
strongly
attractive potential. This implies a
rapid
transition from a repulsive ${\bar K}N$ interaction to an
attractive one as density increases and a microscopic
study of the
$K^-$ properties in the medium cannot be done in terms of the
simple $\bar{T}\rho$ or impulse approximation. It is therefore
necessary to consider
the density dependence of the in-medium ${\bar K}N$
interaction, $T_{\rm eff}(\rho)$.

One source of density dependence comes from the
Pauli principle,
which prevents the scattering to intermediate nucleon states
below the
Fermi momentum, $p_F$.
To incorporate this effect in the channels having an intermediate
nucleon state, one must replace the free nucleon
propagator by the in-medium one in the loop integral of Eq.
(\ref{eq:gprop}).
The evaluation of the $K^-$ self-energy will require the knowledge
of the $K^-N$ $T$-matrix at momenta $P=p_N + p_K$ in the nuclear
lab frame, where $p_N$ is a nucleon momentum from the Fermi sea and
$p_K$ is the momentum of the $K^-$. The loop integral of Eq.
(\ref{eq:gprop}) used in the description of the $K^-N$ scattering
data is evaluated and regularized by means of a cut-off in the
center-of-mass frame. In order to be able to use the same formalism
we boost the $K^-N$ system to its center-of-mass frame and then
evaluate the loop function. However, since Pauli blocking is most
easily implemented in the lab frame, we express the occupation
number in terms of the momentum of the 
intermediate nucleon in the laboratory,
$\vec{q}_{\rm lab}$. Hence, we have
\begin{eqnarray}
G_{l}(P^0,\vec{P},\rho) &=& i \, \int \frac{d^4 q}{(2 \pi)^4} \,
\frac{M_l}{E_l
(-\vec{q}\,)} \, \left\{
\frac{1-n(\vec{q}_{\rm lab})}{\sqrt{s} - q^0 - E_l (-\vec{q}\,) +
i \epsilon} +
\frac{n(\vec{q}_{\rm lab})}{\sqrt{s} - q^0 - E_l (-\vec{q}\,) - i
\epsilon} \right\} \nonumber \\
&\times&
\frac{1}{q^2 - m^2_l + i \epsilon}  \ ,
\label{eq:gpauli0}
\end{eqnarray}
where $(P_0,\vec{P}\,)$ is the total four-momentum in the lab
frame, $s=(P^0)^2-\vec{P}\,^2$ and 
\begin{equation}
\vec{q}_{\rm lab}  =  \left[ - \left(\frac{P^0}{\sqrt{s}} -1
\right) \frac{\vec{P}\vec{q}}{\mid \vec{P} \mid^2} +
\frac{\sqrt{s}-q^0}{\sqrt{s}} \right] \vec{P} - \vec{q} 
\label{eq:lorentz}
\end{equation}
is the nucleon momentum in the lab frame corresponding to a
momentum $-\vec{q}$ in the center-of-mass frame.

As one can see from Eq. (\ref{eq:gpauli0}), the Pauli blocking
corrections are only operative at the nucleon pole. Indeed, by
collecting the terms proportional to $n(\vec{q}_{\rm lab})$ in the
nucleon propagator one finds $2\pi i n(\vec{q}_{\rm lab})
\delta(\sqrt{s}-q^0-E_l(-\vec{q}\,))$, which sets the value of
$q^0$ to be used in Eq. (\ref{eq:lorentz}). This allows one to
perform the integration over $q^0$ in Eq. (\ref{eq:gpauli0})
analytically resulting in 
\begin{eqnarray}
G_{l}(P^0,\vec{P},\rho)
&=& \int_{\mid {\vec q} \mid < q_{\rm max}} \, \frac{d^3 q}{(2
\pi)^3} \,
\frac{1}{2 \omega_l (\vec q\,)}
\,
\frac{M_l}{E_l (-\vec{q}\,)} \,
\left\{
\frac{1 - n(\vec{q}_{\rm lab})}{\sqrt{s}- \omega_l (\vec{q}\,) -
E_l (-\vec{q}\,) + i
\epsilon} \right. \nonumber \\
& &
\phantom{\int_{\mid {\vec q} \mid < q_{\rm max}} \, \frac{d^3
q}{(2
\pi)^3} \,
\frac{1}{2 \omega_l (\vec q\,)}
\,
\frac{M_l}{E_l (-\vec{q}\,)}
}
+ \left. \frac{n(\vec{q}_{\rm lab})}{\sqrt{s} + \omega_l
(\vec{q}\,) -
E_l
(-\vec{q}\,) - i \epsilon} \right\} \ ,
\label{eq:gpauli}
\end{eqnarray}
In practice, we have checked that the value $\sqrt{s}-q^0=E(-
\vec{q}\,)$ to be used in Eq. (\ref{eq:lorentz}) can be replaced,
with negligible changes, by the energy of the nucleon when both the
meson and the baryon are placed on-shell
\begin{equation}
\sqrt{s}-q^0=\frac{s+M_l^2-m_l^2}{2\sqrt{s}} \ .
\end{equation}
The second term on the right hand side of Eq. (\ref{eq:gpauli}),
related to the hole part of the nucleon propagator, is purely
real for the kinematical conditions studied in this work, since
one always finds $\sqrt{s} + \omega(\vec{q}\,) - E_l(-\vec{q}\,) >
0$.

All works studying the Pauli blocking effects on the in-medium
$\bar{K}N$ interaction \cite{Koch94,WKW96,Waas97,Lutz98} have
neglected the hole term of Eq. (\ref{eq:gpauli}), a
simplification that is well justified for low densities
since the phase space for holes is much reduced with
respect to that for particles. However, this effect can become
important if one wants to find the properties of the $\bar{K}$ at
high densities as those achieved in heavy ion reactions or
neutron stars.

Another source of density dependence is related to the fact that
all the mesons and baryons participating in the intermediate
loops interact with the
nucleons of the Fermi sea and, as a consequence, feel a binding
potential which changes the threshold energy of the different
channels.

The binding effects on the baryons are taken within a
mean-field approach consisting in adding, to the single particle
energies in Eqs. (\ref{eq:gprop}) and (\ref{eq:gpauli}), a
momentum-independent potential of the type $U=U_0
\rho/\rho_0$, where $\rho_0=0.17$ fm$^{-3}$ is normal nuclear
matter density. In the case of nucleons a reasonable value is
$U_0=-50$ MeV as
implied by numerous nuclear structure data. On the other hand,
since the experimental binding energy of a $\Lambda$ particle in
hypernuclei extrapolates to about 30 MeV for the very heavy
systems \cite{moto90}, we take the potential depth of the
$\Lambda$
to be $U_0=-30$ MeV. Finally, although there is very little
information on
the $\Sigma$ single particle states, the recent measurement of
the $\Sigma$ binding energy of about 4 MeV in $^4_\Sigma$He
\cite{Nagae98} indicates that the $\Sigma$ can certainly be bound
in the nucleus. Moreover, analysis of $\Sigma$-atom data by Batty
et al \cite{Batty78} and from Ref. \cite{oset90} found the data
to be compatible with a potential of the type $U_0\rho/\rho_0$ with
$U_0$ about $-25$ to $-30$ MeV. Based on this evidence,
we take $U_0=-30$ MeV for the $\Sigma$ potential.

The nuclear medium effects on the mesons will be included through
the corresponding self-energy.
We will only consider the dressing of the
${\bar K}$ and $\pi$ mesons since the
$\eta$ and $K$ mesons appear in intermediate states that lie
quite far above the $K^- p$ threshold. In the next section, we
show how the $\bar{K}$ and $\pi$ self-energies are constructed.
Here, we will focus on how the dressed meson propagator is
incorporated into the scheme. To this end, we write
the meson propagator ($i=\bar{K},\pi$)
\begin{equation}
D_i(q^0,\vec{q},\rho) = \frac{1}{(q^0)^2-{\vec q\,}^2 - m_i^2 -
\Pi_i(q^0,\vec{q},\rho)}
\end{equation}
in the Lehmann representation
\begin{eqnarray}
D_i(q^0,\vec{q},\rho) &=& -\frac{1}{\pi}\int_0^\infty d\omega
\frac{{\rm
Im} D_i(\omega,\vec{q},\rho)}{q^0-\omega + i\epsilon} \,  +
\frac{1}{\pi}\int_0^\infty d\omega \frac{{\rm Im}
D_i(\omega,\vec{q},\rho)}{q^0 + \omega - i\epsilon} \nonumber \\
&=&  \int_0^\infty d\omega \, 2\omega  \, \frac{S_i(\omega,
{\vec q},\rho)}{(q^0)^2 - \omega^2 + i\epsilon}  \ ,
\label{eq:dressed}
\end{eqnarray}
where the spectral
density, defined as
\begin{equation}
S_i(\omega,{\vec
q},\rho)= -\frac{1}{\pi} {\rm Im}\, D_i(\omega,{\vec q},\rho) =
-\frac{1}{\pi}\frac{{\rm Im} \Pi_i(\omega,\vec{q},\rho)}
{\mid \omega^2-\vec{q}\,^2-m_i^2-
\Pi_i(\omega,\vec{q},\rho) \mid^2} \ ,
\end{equation}
has been introduced.

The free meson propagator in Eq.~(\ref{eq:gpauli}) must now be
replaced by the dressed one written as in Eq.~(\ref{eq:dressed}).
While the $q^0$ integral proceeds as
before, there now appears an additional integration over the
variable $\omega$ running over all possible excited states to which
a meson of momentum $\vec{q}$ can couple. The new meson-baryon
loop integral is ($l=\bar{K},{\pi}$)
\begin{eqnarray}
G_l(P^0,\vec{P},\rho)&= &
\int_{\mid
{\vec
q} \mid < q_{max}} \frac{d^3 q}{(2 \pi)^3}
\frac{M_l}{E_l (-\vec{q}\,)}
\int_0^\infty d\omega
 S_l(\omega,{\vec q},\rho) \nonumber
\\
& \times & \left\{
\frac{1-n(\vec{q}_{\rm lab})}{\sqrt{s}- \omega
- E_l (-\vec{q}\,)
+ i \epsilon} +
\frac{n(\vec{q}_{\rm lab})}
{\sqrt{s} + \omega - E_l(-\vec{q}\,) } \right\} \ ,
\label{eq:gmed}
\end{eqnarray}
where $n(\vec{q}_{\rm lab})=0$ for the intermediate states
involving
hyperons ($\pi \Lambda$ and $\pi \Sigma$).

If only Pauli blocking effects are considered, the effective
interaction $T_{\rm eff}(P^0,\vec{P},\rho)$ is
obtained by solving the coupled-channel Bethe-Salpeter
equation using the meson-baryon propagator of Eq. (\ref{eq:gpauli})
for the
intermediate loops involving nucleons and that of Eq.
(\ref{eq:gprop}) otherwise.
If one also incorporates the dressing of the mesons, one must
solve the Bethe-Salpeter equation using the meson-baryon
propagator of Eq. (\ref{eq:gmed}).

\section{Meson self-energies}
\label{sec:self}

The model discussed in Sect. \ref{sec:free} for the $\bar{K}N$
interaction gives rise to a s-wave $\bar{K}$ self-energy (${\bar
K}=K^-$ or $\bar{K}^0$)
\begin{equation}
\Pi^s_{\bar{K}}(q^0,{\vec q},\rho)=2\int \frac{d^3p}{(2\pi)^3}
n(\vec{p}) \left[ T_{\rm eff}^{\bar{K}
p}(P^0,\vec{P},\rho) +
T_{\rm eff}^{\bar{K} n}(P^0,\vec{P},\rho) \right] \ ,
\label{eq:selfka}
\end{equation}
which is obtained by summing the in-medium ${\bar K}N$
interaction, $T_{\rm eff}^\alpha$ ($\alpha={\bar K} p, {\bar
K}n$), over the nucleons in the Fermi sea. The values
$(q^0,\vec{q}\,)$ stand now for the energy and momentum of the
$\bar{K}$ in the lab frame.
Note that a self-consistent approach is required since one
calculates the ${\bar K}$ self-energy from the effective
interaction $T_{\rm eff}$ which uses ${\bar K}$ propagators which
themselves include the self-energy being calculated.

We also include a p-wave contribution to the ${\bar K}$ 
self-energy coming from the coupling of the ${\bar K}$ meson to
hyperon particle-nucleon hole ($YN^{-1}$) excitations.
The $K^-$ meson can couple to $p\Lambda$, $p\Sigma^0$ or
$n\Sigma^-$ and the $\bar{K}^0$ to $n\Lambda$, $n\Sigma^0$ or
$p\Sigma^-$. The vertices $MBB^\prime$ are easily derived from
the $D$ and $F$ terms of Eq.~(1), expanding $U$ up to one meson
field. Using a non-relativistic reduction of the
$\gamma^\mu\gamma^5$ matrix, one finds
\begin{equation}
-i t_{MBB^\prime}=D_{MBB^\prime} \vec{\sigma}\vec{q} =
\left(\alpha_{MBB^\prime} \frac{D+F}{2f_\pi} +
\beta_{MBB^\prime} \frac{D-F}{2f_\pi}\right) \vec{\sigma}\vec{q}
\ ,
\label{eq:coupl}
\end{equation}
where $\vec{q}$ is the kaon momentum. We take $D+F=g_A=1.257$ and
$D-
F=0.33$. The coefficients $\alpha_{MBB^\prime}$ and
$\beta_{MBB^\prime}$ are listed in Table \ref{tab:2}.

The p-wave self-energy for the $K^-$ meson reads
\begin{equation}
\Pi_{K^-}(q^0,\vec{q},\rho_n,\rho_p)=\vec{q}\,^2 (D_{K^-p\Lambda}^2
U_\Lambda(q^0,\vec{q},\rho_p) +
D_{K^-p\Sigma^0}^2  U_\Sigma(q^0,\vec{q},\rho_p) +
D_{K^-n\Sigma^-}^2 U_\Sigma(q^0,\vec{q},\rho_n) ) \ ,
\label{eq:kpwave}
\end{equation}
where $U_Y(q^0,\vec{q},\rho_i)$ is the generalized Lindhard
function for different particle and hole masses in a medium of
density $\rho_i$, with $\rho_i=\rho_n$ or $\rho_p$. As shown in
the appendix of Ref.~\cite{oset90}, it reads
\begin{equation}
U_Y(q^0,\vec{q},\rho_i) = {\rm Re}\, U_Y(q^0,\vec{q},\rho_i) + i
{\rm Im}\, U_Y(q^0,\vec{q},\rho_i) \ ,
\end{equation}
where
\begin{eqnarray}
{\rm Re}\, U_Y(q^0,\vec{q},\rho_i) &=&
\frac{3}{2}\rho_i \frac{M_Y}{q p_F^i} \left\{ z + \frac{1}{2}(1-
z^2) \ln \frac{\mid z + 1\mid}{\mid z - 1 \mid} \right\}
\nonumber \\
{\rm Im}\, U_Y(q^0,\vec{q},\rho_i) &=& -\pi \frac{3}{4} \rho_i
\frac{M_Y}{q p_F^i} \left\{ (1-z^2) \theta(1-\mid z \mid)
\right\} \ ,
\end{eqnarray}
and
\begin{equation}
z=\left( q^0 - \frac{q^2}{2M_Y} - (M_Y-M) \right) \frac{M_Y}{q
p_F^i} \nonumber
\end{equation}

An analogous expression is found for the $\bar{K}^0$ meson. In
symmetric nuclear matter, as is considered here, both
self-energies are the same and reduce to
\begin{equation}
\Pi^{p}_{\bar{K}}(q^0,\vec{q},\rho) =
\frac{1}{2}
\left(\frac{g_{\scriptscriptstyle N\Lambda
K}}{2 M}\right)^2 {\vec q}\,^2 U_\Lambda(q^0,\vec{q},\rho) +
\frac{3}{2} \left(\frac{g_{\scriptscriptstyle N\Sigma K}}{2
M}\right)^2 {\vec q}\,^2 U_\Sigma(q^0,\vec{q},\rho)  \ ,
\label{eq:selfkap}
\end{equation}
where $\displaystyle\frac{g_{\scriptscriptstyle N\Lambda K}}{2
M}=D_{K^-p\Lambda}=D_{\bar{K}^0 n \Lambda}$ and
$\displaystyle\frac{g_{\scriptscriptstyle N\Sigma K}}{2 M}=D_{K^-
p\Sigma^0}$ are the generic $KN\Lambda$ and $KN\Sigma$ coupling
constants, respectively.
This p-wave contribution to the ${\bar K}$ self-energy can become
important for large momentum values.

The solution of the new
$K^-$ dispersion relation
\begin{equation}
\omega^2 = \vec{q}\,^2 + m_K^2 + {\rm Re\, } \Pi_{\bar K}(\omega,{\vec
q},\rho) \ ,
\label{eq:disp}
\end{equation}
with $\Pi_{\bar K}(\omega,{\vec q},\rho)=
\Pi^s_{\bar K}(\omega,{\vec
q},\rho)+\Pi^p_{\bar K}(\omega,{\vec
q},\rho)$,
determines the effective mass, $m^*_K={\rm
Re}\, \omega({\vec q}=0)$, and decay width, $\Gamma=-2\,{\rm Im}
\,\omega({\vec q}=0)$, of the $K^-$ meson in the medium.

For the pion self-energy we take that of Ref.
\cite{ramos94} which consists of a small momentum independent
s-wave part, $\Pi_\pi^s(\rho)$, plus a p-wave part,
$\Pi_\pi^p(q^0,\vec{q},\rho)$. The latter, constructed by
allowing the pion to couple to particle-hole ($1p1h$),
$\Delta$-hole ($\Delta h$) and two-particle-hole ($2p2h$)
excitations, can be written as
\begin{equation}
\Pi_\pi^p(q^0,\vec{q},\rho)= \left(\frac{g_{\pi NN}}{2M} \right)^2
\vec{q}\,^2 F^2(q) (U_N(q^0,\vec{q},\rho) +
U_\Delta(q^0,\vec{q},\rho) + U_{2p2h}(q^0,\vec{q},\rho)) \ ,
\label{eq:pself}
\end{equation}
where $F(q)$ is a monopole form factor with cut-off $\Lambda_\pi=1.2$
GeV, $U_N$ ($U_\Delta$) is the Lindhard function for $1p1h$
($\Delta h$) excitations and $U_{2p2h}$ is a phenomenological function
accounting for $2p2h$ excitations. The $U_{2p2h}$ contribution at
$(q^0,\vec{q},\rho)=(m_\pi,\vec{0},0.75\rho_0)$ was obtained
from fits to pionic atom data \cite{meirav89}.
The extension to kinematical regions away from pionic atoms was
done
by multiplying the imaginary part by the available phase space of
$2p2h$ states for the given $q^0$, $\vec{q}$ values at density
$\rho$.
The explicit expression can be found in Ref. \cite{ramos94}.

The pion self-energy is further modified to include the effect of
nuclear short-range correlations. This is accomplished by
replacing the self-energy of
Eq.~(\ref{eq:pself}) by
\begin{equation}
\Pi^p_\pi(q^0,\vec{q},\rho)
=\left(\frac{g_{\pi NN}}{2M} \right)^2 \vec{q}\,^2
 F^2(q) \frac{\Pi^{(0)}_\pi(q^0,\vec{q},\rho)}
 {1- \left(\frac{g_{\pi NN}}{2M} \right)^2 g^\prime(q)
 \Pi^{(0)}_\pi(q^0,\vec{q},\rho) } \ ,
\end{equation}
where $\Pi^{(0)}_\pi$ is the sum of the $U_{1p1h}$, $U_{\Delta
h}$ and $U_{2p2h}$ functions, and
$g^\prime$ is the usual Landau-Migdal parameter, which is taken
slightly momentum dependent as in Ref. \cite{ramos94}.

We have also considered the effects of including form factors at
the $\bar{K}NY$ vertices of Eq. (\ref{eq:kpwave}). If a monopole
form factor with $\Lambda_K=1.3$ GeV normalized to 1 as $q^\mu \to 0$
is used, the changes observed are very small, at the level of 2\%.
Since the value of $\Lambda_K$ for kaons is not so well known and the
changes induced by the form factor are so small, we simply ignore
it in the results presented below.

\section{Results}
\label{sec:results}

In this section we discuss the results of our calculations. To
facilitate the comparison with other works we will distinguish
three types of approximations, all of which include the binding
effects on the baryons:
\begin{itemize}
\item[a)]
{\it Pauli}: The blocking of
intermediate
nucleon states below the Fermi momentum $p_F$ is taken into
account, but the mesons
propagate as in free space. This approach uses the meson-baryon
propagator of Eq. (\ref{eq:gpauli}) for the intermediate states
involving nucleons and has been the standard medium effect
studied in the literature \cite{Koch94,WKW96,Waas97}.
\item[b)]
{\it In-medium kaons}: apart from Pauli blocking, the dressed
${\bar
K}$ propagator is included in the loops in a self-consistent
manner.
This approach, which uses Eq. (\ref{eq:gmed}) for the intermediate
${\bar K}N$ states, has recently
been used in the study of the $K^-$ properties in the medium
\cite{Lutz98}.
\item[c)]
{\it In-medium pions and kaons}: the dressing of the pions is
also
considered and, therefore, Eq. (\ref{eq:gmed}) is used for the
intermediate channels involving both ${\bar K}$ and $\pi$ mesons.
This is the main novelty of the present work, apart from
considering also the baryon potentials.
\end{itemize}

\begin{figure}
       \setlength{\unitlength}{1mm}
       \begin{picture}(100,150)
       \put(25,10){\epsfxsize=12cm \epsfbox{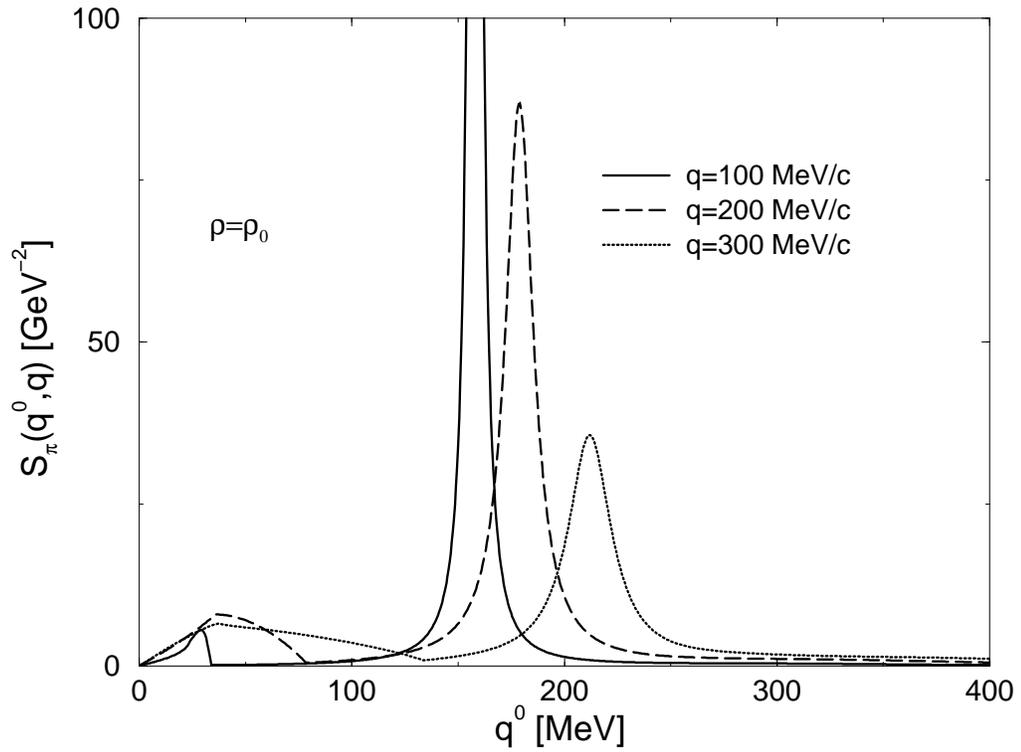}}
       \end{picture}
\caption{
Pion spectral density at normal nuclear matter density as a
function
of energy for several pion momenta: $q=100$ MeV/c (solid line),
$q=200$ MeV/c 
(dashed line) and $q=300$ MeV/c (dotted line).
\label{fig:specpi}}
\end{figure}

To assess the importance of dressing the mesons, we show first
their spectral density. In Fig. \ref{fig:specpi} the spectral
density of the $\pi$ mesons in nuclear matter at density
$\rho=\rho_0$ is shown as a function of energy for several
momenta. Note that the scale has been adapted to the two higher
momenta such that the structure of the spectral density is more
clearly seen. The peak value of the $q=100$ MeV/c spectral
function is about 375 GeV$^{-2}$. As the momentum increases the
position of the peak moves to higher energies. However, it
increasingly deviates from the energy $\sqrt{\vec{q}\,^2 +
m_\pi^2}$ at which a delta-type spectral density would be located
if the dressing of the $\pi$ meson was ignored. The in-medium peak
positions are
at 158, 179 and 212 MeV  for $q=100$, 200 and $300$ MeV/c,
respectively,
while the corresponding free values are 172, 244 and 331 MeV.
We also observe, to the left of the peaks, the typical structure
of the $1p1h$ excitations, which spreads the pionic strength over
a wider energy range.

\begin{figure}
       \setlength{\unitlength}{1mm}
       \begin{picture}(100,180)
       \put(25,10){\epsfxsize=12cm \epsfbox{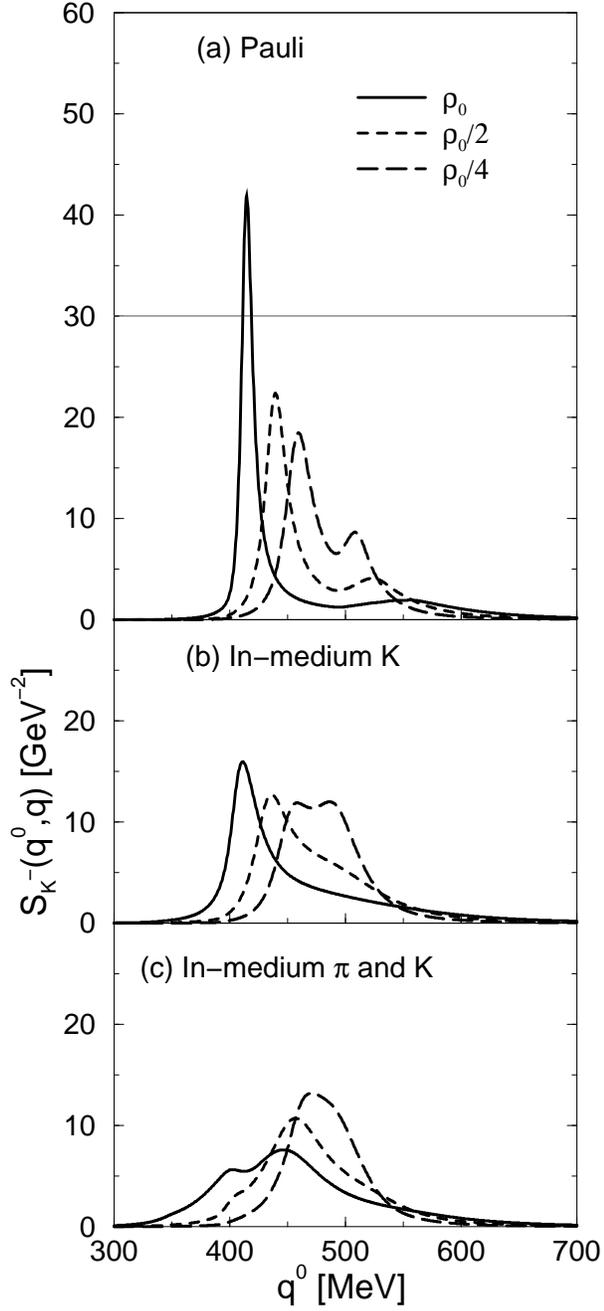}}
       \end{picture}
\caption{
$K^-$ spectral density for zero momentum as a function of energy at
several
densities: $\rho_0$ (solid line), $\rho_0/2$
(short-dashed line)
and $\rho_0/4$ (long-dashed line).
Results are 
shown for the three approximations discussed in the text: a) {\it
Pauli} (top panels), 
b) {\it In-medium kaons} (middle panels) and c) {\it In-medium
pions and kaons} (bottom
panels).
\label{fig:specka}}
\end{figure}

The spectral funcion of a $K^-$ meson of zero momentum is
shown in Fig. \ref{fig:specka} as a function of energy for the
three types of approximations used in this work. The solid line
shows results for nuclear matter density $\rho=\rho_0$, the
short-dashed line for $\rho_0/2$ and the long-dashed line for
$\rho_0/4$.

The spectral functions shown in the upper panel correspond to
the approximation in which only Pauli blocking effects have been
considered. For $\rho_0/4$ a two-mode excitation is clearly
visible. The left peak corresponds to the $K^-$ pole branch,
appearing at an energy smaller than its mass, $m_K$, due to the
medium effects which, as we will see, are already attractive at
this density. The upper peak corresponds to the
$\Lambda(1405)$-hole
state, which is located above $m_K$ because of the shifting of
the $\Lambda(1405)$ resonance to energies above the $K^-p$
threshold due to the restrictions on phase space imposed by Pauli
blocking. As density increases, the $K^-$ feels an enhanced
attraction and the $K^-$ pole peak moves to lower energies. At
the same time the reduced phase space for the intermediate
nucleon states makes the appearance of the resonance less likely.
The $\Lambda(1405)$-hole peak moves to higher energies and loses
strength, a reflection of the tendency of the $\Lambda(1405)$
to dissolve in the dense nuclear medium.
These results are in complete agreement with those obtained by
Waas and Weise \cite{Waas97}.

When the ${\bar K}$ spectral function displayed in Fig.
\ref{fig:specka}
is insterted in the coupled-channel Bethe-Salpeter equation,
the intermediate ${\bar K}N$ states are more spread out over
energies. As a consequence, the resulting in-medium interaction,
$T_{\rm eff}$, and the new self-energy calculated from Eq.
(\ref{eq:selfka}),  become softened and distributed over a wider
range of energies, affecting in turn the distribution of the
${\bar K}$
spectral strength. After a few iterations of this process, self-
consistency is achieved and the resulting $K^-$ spectral function
is displayed in the middle panel of Fig. \ref{fig:specka}. The
two peak mode is barely visible now. As was noted by Lutz
\cite{Lutz98}, including the in-medium attraction felt by the
$K^-$ through the use of the dressed propagator lowers the
threshold for the ${\bar K}N$ states that had been increased by
the Pauli blocking on the nucleon. This has a compensatory effect
and the resonance barely moves with respect to its free space
value.
The $K^-$ pole peak appears at similar or slightly
smaller energies, but its width is larger, due to the
strength of the intermediate ${\bar K}N$ states being distributed
over a wider region of energies.
Therefore the $K^-$ pole and the $\Lambda(1405)$-hole branches
merge into one another and can hardly be distinguished.

Finally, the self-consistent $K^-$ spectral function when the
pions in the intermediate loops are also dressed is shown
in the bottom panel of Fig. \ref{fig:specka}. The essential
difference with respect to the other two approaches is the fact
that the $\pi Y$ channels ($Y=\Lambda$ or $\Sigma$) start to be
operative at lower energies due to the attraction felt by the
pions in the medium. As seen by the long-dashed line, even at
very small densities one no longer distinguishes the
$\Lambda(1405)$-hole
peak from the $K^-$ pole one. As density increases the attraction
felt by the $K^-$ is much more moderate and the $K^-$ pole peak
appears at
higher energies than in the other two approaches. At
normal nuclear matter density a peak starts to develop to the
left of that corresponding to the $K^-$
pole. As will be shown next, this structure is associated to
a cusp in the imaginary part of the $K^-$ self-energy at an
energy corresponding to the position of the new in-medium $\pi
\Sigma$ threshold.

\begin{figure}
       \setlength{\unitlength}{1mm}
       \begin{picture}(100,180)
       \put(25,10){\epsfxsize=12cm \epsfbox{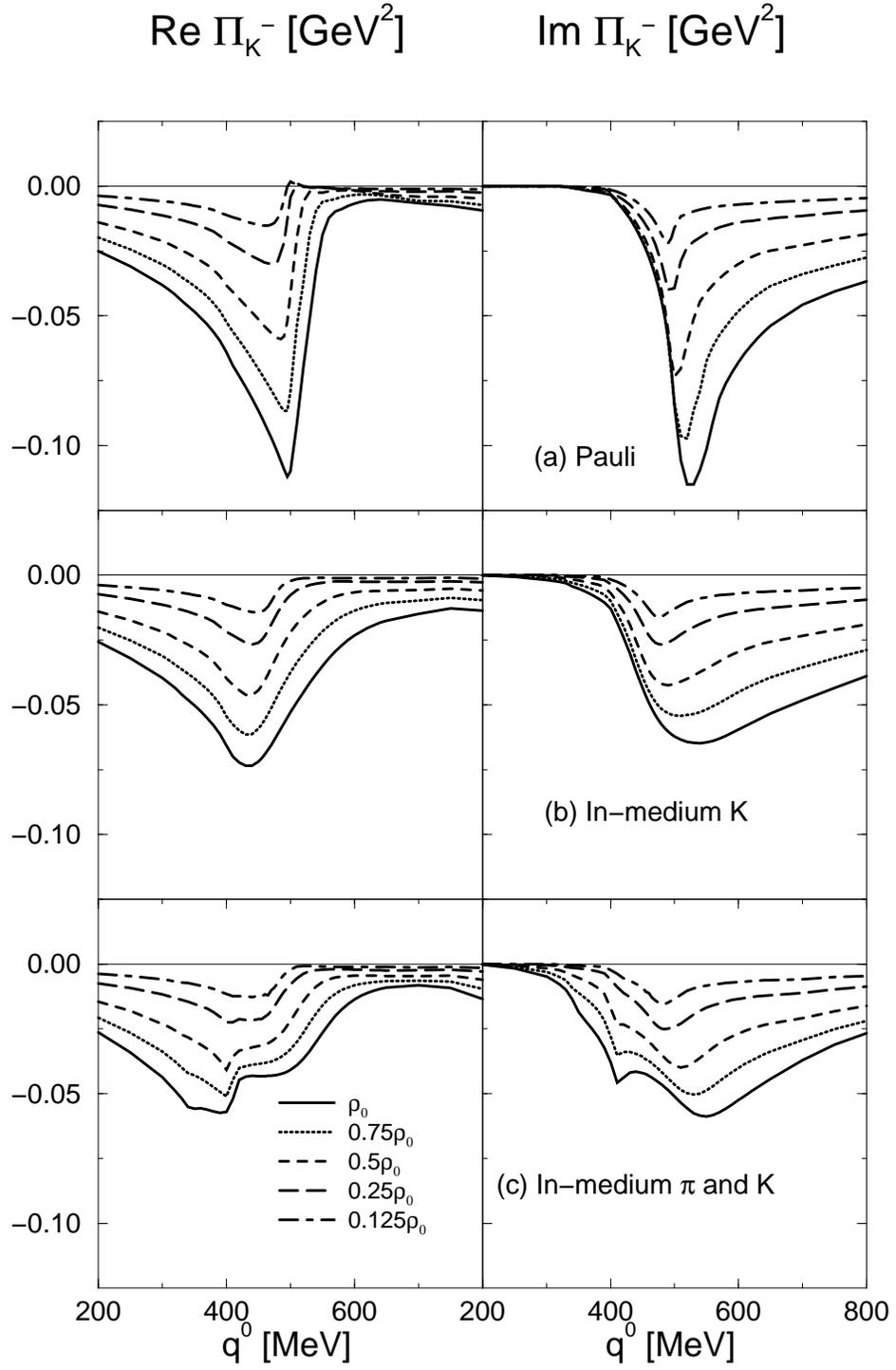}}
       \end{picture}
\caption{
Real (left) and imaginary (right) parts of the zero momentum $K^-$
self-energy 
as a function of energy, for several nuclear matter densities. 
Results are 
shown for the three approximations discussed in the text: a) {\it
Pauli} (top panels), 
b){\it In-medium kaons} (middle panels) and c) {\it In-medium
pions and kaons} (bottom
panels).
\label{fig:selftot}}
\end{figure}

The $K^-$ self-energy at zero momentum, $\Pi_{K^-}(q^0,0,\rho)$,
is displayed
in Fig. \ref{fig:selftot} as a function of the $K^-$ energy for
several densities and for the three approximations discussed
here. The graphs on the left show the real part and those on the
right the imaginary part. We have seen that the ${\bar K}$
spectral strength is
distributed over a wider range of energies and
incorporates the attractive medium effects felt by the ${\bar K}$.
This is why the self-energy shown in the middle panels ({\it
In-medium kaons}) is smoother and wider than that
displayed on the top ({\it Pauli})
ones. Note that ${\rm Im}\,\Pi_{K^-}$ in the middle panel is
non-zero at energies as low as 300 MeV. 
This corresponds to the $N(\Lambda h)$ and $N(\Sigma h)$ many-body
channels that are opened at these energies and which are
incoporated in the calculation via the $K^-$ p-wave self-energy. In
the lower panels, where the pions are also dressed, we see that
$Im \Pi_{K^-}$ has bigger strength at these low energies. This can
be understood in the following way: the $\pi\Sigma$ channel is one
of the important building blocks of the $\Lambda(1405)$ strength
below the $\bar{K}N$ threshold. When the pions are dressed in the
medium they couple strongly to $1p1h$ and $2p2h$ components, so
that the $K^-$ is effectively coupling to intermediate
$1p1h\Sigma$, $2p2h\Sigma$ states with lower threshold than the
$\pi\Sigma$ channel. Actually, the opening of this latter channel
on top of the already existing $1p1h\Sigma$ and $2p2h\Sigma$ is
visible through a cusp in ${\rm Im}\, \Pi_{K^-}$ around $q^0 \simeq
400$ MeV.

The effect on the $K^-$ self-energy of the baryon binding
potentials is basically a shift in the scale of energies,
corresponding to the difference of potentials between the nucleon
and the $\Sigma$ hyperon, which amounts to 20 MeV at $\rho=\rho_0$.

\begin{figure}
       \setlength{\unitlength}{1mm}
       \begin{picture}(100,180)
       \put(25,10){\epsfxsize=12cm \epsfbox{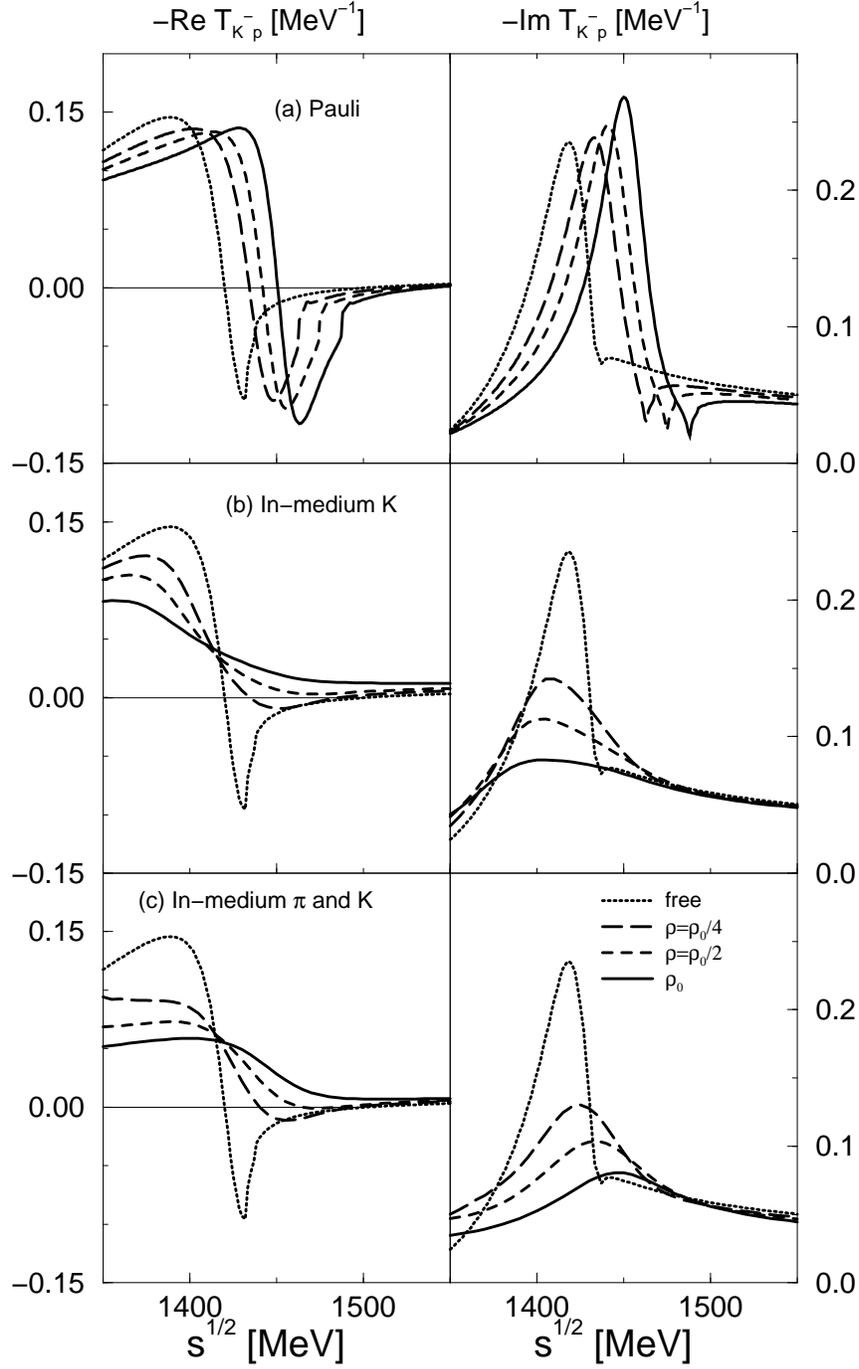}}
       \end{picture}
\caption{
Real (left) and imaginary (right) parts of the in-medium $K^- p$
scattering amplitude
as a function of the invariant energy
$\sqrt{s}$ for $\mid \vec{p}_{K^-} + \vec{p}_N \mid = 0$ and
several densities. Results are
shown for the three approximations discussed in the text: a) {\it
Pauli} (top panels), 
b) {\it In-medium kaons} (middle panels) and c) {\it In-medium
pions and kaons} (bottom
panels).
\label{fig:ampltot}}
\end{figure}

The scattering properties of a ${\bar K}$ in the nuclear medium
will depend on the characteristics of the in-medium ${\bar K}N$
scattering amplitude. In particular, it is interesting to see if
the resonant shape of the free $K^- p$ scattering amplitude
remains when the medium effects are incorporated. Fig.
\ref{fig:ampltot} shows the real part (on the left) and imaginary
part (on the right) of the $K^-p \to K^-p$ scattering amplitude
for a total momentum $\mid \vec{p}_K + \vec{p}_N \mid =0$ as a
function of $\sqrt{s}$. Results are shown for several densities
and for the three types of approximations.
On the top panels we can see how, as density increases, the
Pauli blocking effects merely shift the resonance to higher
energies, hence changing from being 27 MeV below
the $K^- p$ threshold to being
above it from a certain density on. Note, however, that the
resonance still appears, for each density, below the new
threshold imposed by the Pauli principle, i.e. $m_K + M +
p_F^2/2 M$. On the other hand, the resonance shape remains
pretty much unaltered.
When the dressing of the ${\bar K}$ meson is also incorporated,
we observe how the resonant shape smears out quite fast with
increasing density. Moreover, the resonance does not move up to
higher energies but stays close to the free space value due to a
cancellation between the repulsive Pauli blocking effects and the
attractive medium effects on the ${\bar K}$. These results are in
qualitative agreement with the self-consistent calculation of Lutz
\cite{Lutz98}, although we seem to observe a faster dilution of
the resonance with the medium density.
Finally, when the in-medium dressing of the pions are also
included we obtain a quite similar pattern. The resonance
dissolves very fast although it shows a tendency to move to higher
energies.

\begin{figure}
       \setlength{\unitlength}{1mm}
       \begin{picture}(100,180)
       \put(25,10){\epsfxsize=12cm \epsfbox{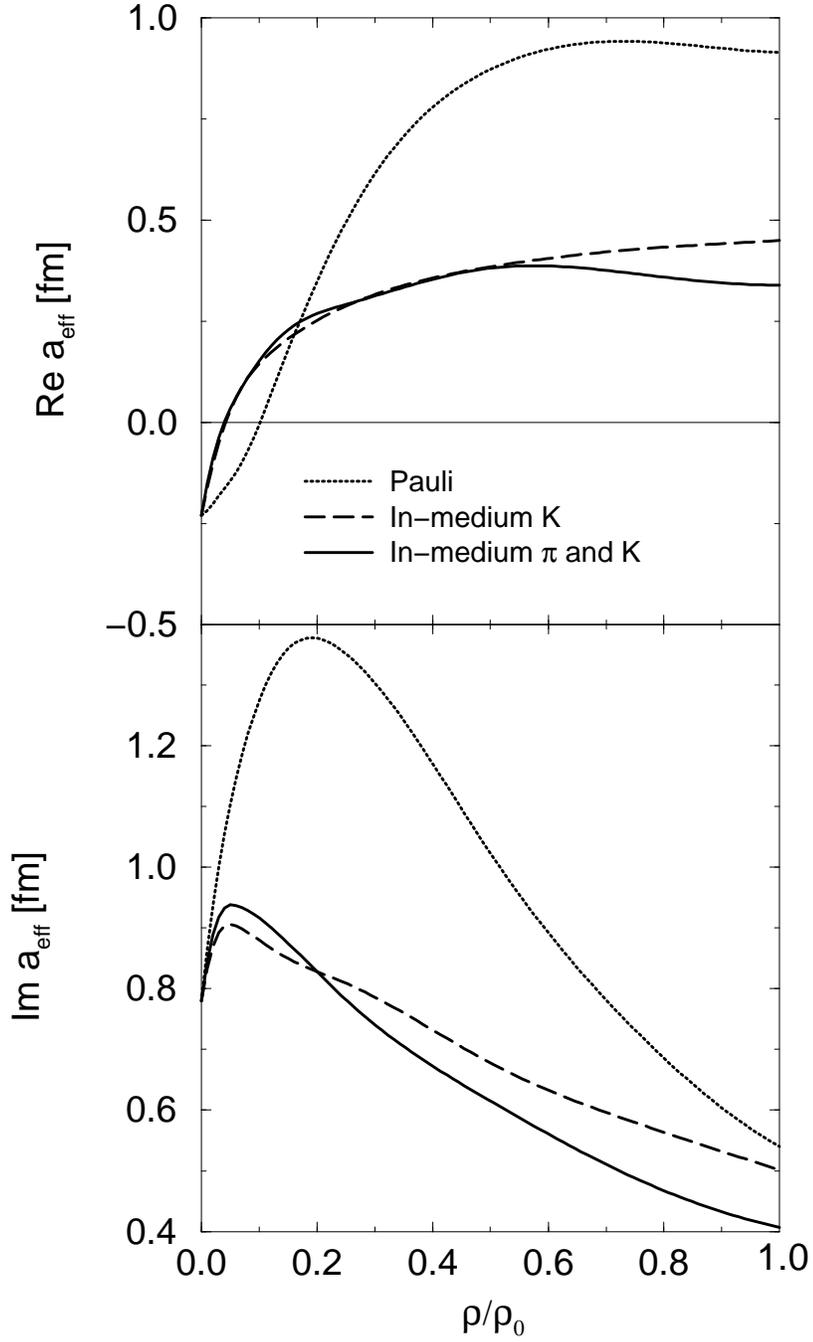}}
       \end{picture}
\caption{
Real (top) and imaginary (bottom) parts of the in-medium $K^-N$
scattering length
as a function of density for the
three approximations discussed in the text: {\it Pauli} (dotted
lines), 
{\it In-medium kaons} (dashed lines) and {\it In-medium pions and
kaons} (solid
lines).
\label{fig:scatlen}}
\end{figure}

The isospin averaged in-medium scattering length defined as
\begin{equation}
a_{\rm eff}= -\frac{1}{4\pi} \frac{M}{m_K + <E_N>}
\frac{\Pi_{\bar{K}}(m_K,\vec{q}=0,\rho)}{\rho} \ ,
\end{equation}
with
$<E_N>=M+U_N(\rho)+\displaystyle\frac{3}{5}\left(\displaystyle\frac{ p_F
^2}{2 M}\right)$,
is shown if Fig. \ref{fig:scatlen}. The change of ${\rm Re}\,
a_{\rm eff}$ from negative to positive values indicates the
transition from a repulsive interaction in free space to an
attractive interaction in the medium. As shown by the dotted
line, this transition happens at a density of about $\rho\sim
0.1\rho_0$ when only Pauli effects are considered, in agreement
with what was found in Ref. \cite{WKW96}. However, this
transition occurs at even lower densities ($\rho \sim 0.04
\rho_0$) when one considers the
dressing of the mesons in the description, as seen from the
dashed ({\it In-medium kaons}) and full ({\it In-medium
pions and kaons})
lines. The deviations from the approach including only Pauli
blocking or those dressing the mesons are quite appreaciable over
a wide range of densities.

\begin{figure}
       \setlength{\unitlength}{1mm}
       \begin{picture}(100,100)
       \put(25,10){\epsfxsize=12cm \epsfbox{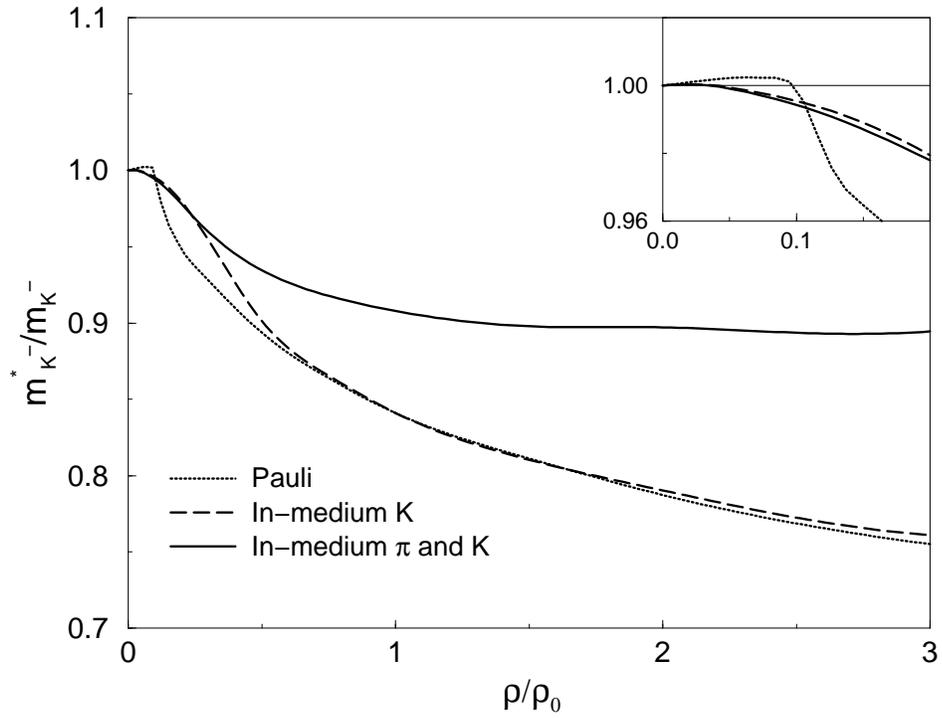}}
       \end{picture}
\caption{
Effective mass of the $K^-$ meson in nuclear matter as a function
of density
for the 
three approximations discussed in the text: {\it Pauli} (dotted
lines), 
{\it In-medium kaons} (dashed lines) and {\it In-medium pions and
kaons} (solid
lines).
\label{fig:emass}}
\end{figure}

The effective $K^-$ mass is shown in Fig. \ref{fig:emass} as a
function of $\rho/\rho_0$ for the different approximations. Note
that the effective mass, defined as the solution of Eq.
(\ref{eq:disp}) for $\vec{q}=0$, gives the position of the $K^-$
pole peak in the
spectral function assuming a weak energy dependence of ${\rm
Im}\, \Pi_{K^-}(q^0,0,\rho)$. In addition, one can only think of
the $K^-
$ as a quasi-meson moving with the modified $m^*_{K^-}$ mass if
the spectral function shows a clear and narrow enough $K^-$ pole
peak, which happens when ${\rm Im}\, \Pi_{K^-}(q^0,0,\rho)$ is
sufficiently small.
By looking at the $K^-$ spectral function of Fig.
\ref{fig:specka} one sees that these conditions are not always
fulfilled, especially when the dressing of the mesons is
incorporated. Keeping this in mind, we can still analyze the
effective mass as a simplified way to assess how important is the
attraction felt by the $K^-$ meson in the medium. Morover, this
analysis will allow to compare our results with those obtained by
other calculations.
As can be seen in the zoomed area on the top right of the figure,
the in-medium $K^-$ mass increases slightly from the free value
but quickly gains attraction as density increases. The transition
from repulsion to attraction occurs around $0.1\rho_0$ for the
{\it Pauli} approximation (dotted line) while the two other
approximations that incorporate the dressing of the ${\bar K}$
(dashed line) and also that of the pion (full line) show
attraction already at around $0.04\rho_0$. However, the density
dependence is stronger with the {\it Pauli} approximation which
gives
rise to a lower value of $m^*_{K^-}$ from $0.11\rho_0$ onwards.
It turns out that, in spite of the essentially different shapes
of the $K^-$ spectral function, the position of the $K^-$
pole is very similar for the {\it Pauli} and the {\it In-medium
kaons}
approximations for densities higher than $\rho_0$. Consequently,
the calculated effective masses decrease similarly. However, when
the dressing of the pions is also included, the attraction is
substantially more moderate and the effective mass seems to level
off at high densities around the value achieved already at $\rho_0$
of 
$m^*_{K^-}=0.9 m_{K}=445$ MeV. This would correspond to a $K^-$
optical
potential of $U_{K^-}={\rm Re}\,
\Pi_{K^-}(m^*_{K^-},\vec{0},\rho_0)/(2m^*_{K^-}) \sim -50$ MeV. The
{\it
Pauli} and {\it In-medium kaons} approximations give rise to a
a more attractive value of the effective mass at $\rho_0$, namely
$m^*_{K^-} = 
415$ MeV which corresponds to a $U_{K^-}=-86$ MeV. It is worth
mentioning that
all the calculations, either microscopic \cite{WKW96,Waas97} or
based on mean
field
theories \cite{Li97,Mao99,Sch97,Tsushi98} predict
effective masses at $\rho_0$ which range between 375 MeV and 425
MeV. The
corresponding $K^-$ potential at the center of the nucleus would
then be in
between $-140$ and $-75$ MeV. Therefore, no theory is able to
explain the large attraction of $-200$
obtained from the best available fit to kaonic atom data
\cite{FGB94}. 
However, we should warn that the data of $K^-$ atoms are only
sensitive to the $K^-$ self-energy at the low effective densities
felt by the kaons that explore basically the nuclear surface. The
values obtained by the fit in Ref. \cite{FGB94} at $\rho=\rho_0$
are an extrapolation of a preassumed functional dependence on
$\rho$. It is possible to obtain equally good fits with other
functionals that give the same value for the low densities explored
by the kaons but different extrapolations at $\rho=\rho_0$. In
fact, quite good fits to the $K^-$ atoms data are also obtained
with the Batty potential of Ref. \cite{batty81} which is of the
order of $-50 \rho/\rho_0$ MeV. 
We also note that the Brueckner-type calculations of Ref.
\cite{alberg76} obtained a shallow $K^-$-nucleus potential, of the
order of $-40$ MeV at the center of $^{12}$C, and predicted
reasonably well the data available at that time.  

With the values obtained here for $m^*_{K^-}$ and similar  ones
that we obtain for neutron matter the phenomenon of $K^-$
condensation appears very unlikely, since
there is not enough attraction for the $K^-$ energy to become
smaller than the electron chemical potential in beta stable
neutron star matter.

\begin{figure}
       \setlength{\unitlength}{1mm}
       \begin{picture}(100,180)
       \put(25,10){\epsfxsize=12cm \epsfbox{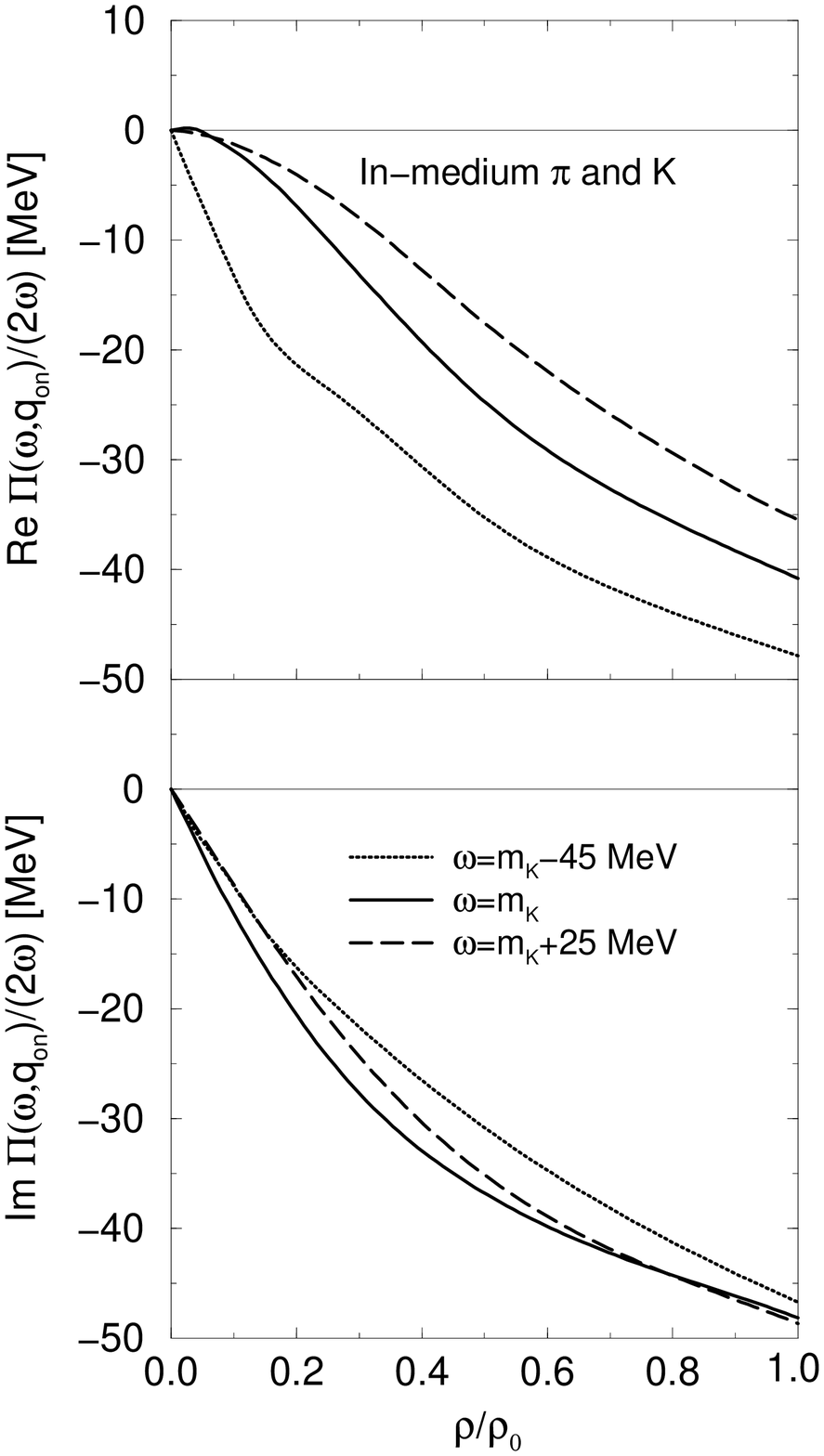}}
       \end{picture}
\caption{
Real (top) and imaginary (bottom) parts of the $K^-$ optical
potential as
a function of density obtained 
from the {\it In-medium pions and kaons} approximation.
Results are shown for three different $K^-$ energies:
$\omega=m_{K}-45$ MeV (dotted lines), $\omega=m_{K}$ (solid lines)
and
$\omega=m_{K}+20$ MeV (dashed lines).
\label{fig:optpotpika}}
\end{figure}

The implications of our results for the in-medium $K^-$
self-energy on kaonic atoms is currently being analyzed
\cite{oku99} and gives results compatible with the existing kaonic
atom data. 
That analysis, based on a local density approximation, requires the
knowledge
of the optical potential,
defined as
\begin{equation}
U_{K^-}(\omega,\rho)=\frac{\Pi_{K^-}(\omega,q_{\rm
on},\rho)}{2\omega} \ ,
\end{equation}
where $q_{\rm on}$ is the on-shell momentum value that fulfills
Eq. (\ref{eq:disp}) for a given value of $\omega$.
The real and imaginary parts of the optical potential for the
{\it In-medium pions and kaons} approximation are shown in Fig.
\ref{fig:optpotpika} for three energy values: $m_{K}-45$ MeV
(dotted line),  $m_{K}$ (solid line) and $m_{K}+25$ MeV
(dashed line). The self-energy shown by the solid lines is the
relevant one for studies of low lying $K^-$ atomic states, while
the results displayed by the dotted line would be the ones to use
for very deeply
bound states in the Pb region. The dashed lines show an example of
the $K^-$
optical potential at positive energies and  would be the one to
consider to treat the $K^-$ distortions in $K^-$-nucleus
scattering reactions around 25 MeV.
While the real part of the optical potential becomes less
attractive with increasing energy, the imaginary part does not
show a too strong energy dependence in the range of energies
explored. This behavior can be easily inferred from the $K^-$
self-energies shown at the bottom of Fig. \ref{fig:selftot}. 


\section{Conclusions}
\label{sec:conclusions}

In this work we have studied the $K^-$ properties in 
nuclear matter using a chiral unitary approach for the $K^- N$
interaction in s-wave that incorporates
the medium effects microscopically. 

Pauli blocking acting on the intermediate
nucleon states modifies the $K^- N$ interaction from that in free
space and 
gives rise to a $K^-$ spectral function that shows two distinctive
peaks at very
low density. The lower one corresponds to the 
the position of the new
$K^-$ pole, located below the $K^-$ mass from already quite low
densities on, which indicates that the
$K^-$ feels an attraction in the medium.
The higher peak is associated to the in-medium 
$\Lambda(1405)$ resonance that appears above the free $K^-
p$ threshold due to the repulsive effect induced by blocking the
intermediate nucleon states below the Fermi momentum. As density
increases, the
$K^-$ feels more attraction and the $\Lambda(1405)$ keeps moving to
higher energies.

When the $K^-$ self-energy is incorporated self-consistently into
the scheme, there is a compensation between the attraction felt by
the $K^-$ and
the repulsive Pauli blocking shift. As a result, the resonance
appears at a
similar location as that in free space. However, its width is much
larger and
it tends to dissolve with increasing density due to a weaker
in-medium  $K^- N$
interaction which now is determined from dressed ${\bar K}$
propagators which
are more spread out over
energies.

In the present work we also include the dressing of the pions in
the
intermediate $\pi\Lambda$, $\pi\Sigma$ states to which the ${\bar
K}N$ system
can couple.
We use a pion self-energy that includes the coupling to $1p1h$,
$\Delta h$ and
$2p2h$ excitations modified by the effect of short-range $NN$
correlations. The
fact that now the pions also have a spreading width makes the
in-medium ${\bar
K}N$ interaction even smoother. The $K^-$ feels less attraction and
the $\Lambda(1405)$
resonance, which is shifted slighty upwards from its free space
position,
dissolves even faster with density.

Our approach gives rise to a $K^-$ self-energy at normal nuclear
density which
has about half the attraction of that obtained with other theories
and
approximation schemes
and would make the phenomenon of kaon
condensation very unlikely.
The self-energy obtained is, however, compatible 
with the existing data on kaonic atoms.

\section*{Acknowledgments}

This work is partially supported
by DGICYT contract numbers PB95-1249 and PB96-0753,
 and by the
EEC-TMR Program under contract No. CT98-0169.


\begin{table}
\caption{$K^-p$ threshold ratios and $K^-N$ scattering lengths}
\begin{tabular}{lcc}
 & This work & Exp.  \\
\hline 
$\gamma=\displaystyle\frac{\Gamma(K^-p\to \pi^+ \Sigma^-)}{
\Gamma(K^-p \to \pi^-\Sigma^+)}$ & 2.32 & $2.36 \pm 0.04$
\cite{To71,No78} \\
 & & \\
$R_c=\displaystyle\frac{\Gamma(K^-p\to {\rm charged\ particles})}{
\Gamma(K^-p \to {\rm all})}$ & 0.627 & $0.664 \pm 0.011$
\cite{To71,No78} \\
 & & \\
$R_n=\displaystyle\frac{\Gamma(K^-p\to \pi^0\Lambda)}{
\Gamma(K^-p \to {\rm neutral\ states})}$ & 0.213 & $0.189\pm0.015$ 
\cite{To71,No78} \\
$a_{K^-p}$ (fm) & $-1.00 + i 0.94$ & $-0.67 + i 0.64$ \cite{Adm81}
\\
                &                & $-0.98$ (from Re($a$))
\cite{Adm81} \\
         &          & $(-0.78\pm0.18) + i(0.49\pm0.37)$
\cite{Iw97}\\
$a_{K^-n}$ (fm) & $0.53 + i 0.62$ & $0.37 + i 0.60$ \cite{Adm81} \\
                &                & $0.54$ (from Re($a$))
\cite{Adm81} 
\end{tabular}
\label{tab:1}
\end{table}

\begin{table}
\caption{SU(3) coupling constants defined in Eq.
(\protect\ref{eq:coupl})}
\begin{tabular}{c|cc|c|cc|}
 &\multicolumn{2}{c|}{$K^-$} & &
  \multicolumn{2}{c|}{$\bar{K}^0$} \\
 & $\alpha_{MBB^\prime}$ & $\beta_{MBB^\prime}$ &
 & $\alpha_{MBB^\prime}$ & $\beta_{MBB^\prime}$ \\
\hline
$p\Lambda$ & $-\frac{2}{\sqrt{3}}$ & $\frac{1}{\sqrt{3}}$  & 
$n\Lambda$ & $-\frac{2}{\sqrt{3}}$ & $\frac{1}{\sqrt{3}}$ \\
$p\Sigma^0$ & $0$ & $1$ & $n\Sigma^0$ & $0$ & $-1$ \\
$n\Sigma^-$ & $0$ & $\sqrt{2}$ & $p\Sigma^+$ & $0$ & $\sqrt{2}$ 
\end{tabular}
\label{tab:2}
\end{table}


\begin{references}
\bibitem{KN86} D.B. Kaplan and A.E. Nelson, Phys. Lett. 
B175 (1986) 57
\bibitem{kaos97} R. Barth et al., Phys. Rev. Lett. 78 (1997)
4007 
\bibitem{cassing97} W. Cassing, E.L. Bratkovskaya, U. Mosel, S.
Teis and A.
Sibirtsev, Nucl. Phys.  A614 (1997) 415; E.L. Bratkovskaya,
W. Cassing and
U. Mosel, Nucl. Phys.  A622 (1997) 593 
\bibitem{Li97} G.Q. Li, C.-H. Lee and G.Brown, Nucl. Phys. 
A625 (1997) 372; {\it ibid}, Phys. Rev. Lett. 79 (1997) 5214 
\bibitem{FGB94} E. Friedman, A. Gal, C.J. Batty, Nucl. Phys. 
A579 (1994) 518
\bibitem{Lee9596} C.-H. Lee, G.E. Brown, D.P. Min and M. Rho,
Nucl. Phys.  A385 (1995) 481;
C.-H. Lee, Phys. Rep. 275 (1996) 255
\bibitem{Kai9597} N. Kaiser, P.B. Siegel and W. Weise, Nucl.
Phys.  A594 (1995) 325; N. Kaiser, T. Waas and W. Weise, Nucl.
Phys. A612 (1997) 297 
\bibitem{oset98} E. Oset and A. Ramos, Nucl. Phys. A635 (1998) 99
\bibitem{Koch94} V. Koch, Phys. Lett. B337 (1994) 7
\bibitem{WKW96} T. Waas, N. Kaiser and W. Weise, Phys. Lett.
B365 (1996) 12; {\it ibid.} B379 (1996) 34
\bibitem{Waas97} T. Waas and W.
Weise, Nucl. Phys. A625 (1997) 287
\bibitem{Lutz98} M. Lutz, Phys. Lett. B426 (1998) 12
\bibitem{alberg76} M. Alberg, E.M. Henley and L. Wilets, Ann. Phys.
96 (1976) 43
\bibitem{Mao99} G. Mao, P. Papazoglou, S. Hofmann, S. Schramm, H.
St\"ocker
and W. Greiner, Phys. Rev. C, in press. {\tt
nucl-th/9806068}
\bibitem{Sch97} J. Schaffner and I.N. Mishustin, Phys. Rev. 
C53 (1996) 1416; \\ J. Schaffner-Bielich, I.N. Mishustin and
J. Bondorf, Nucl. Phys. A625 (1997) 325 
\bibitem{Tsushi98} K. Tsushima, K. Saito, A.W. Thomas and S.V.
Wright, Phys.  Lett. B429 (1998) 239
\bibitem{Ga85} J. Gasser and H. Leutwyler, Nucl. Phys. 
B250 (1985) 465
\bibitem{Pi95} A. Pich, Rep. Prog. Phys.  58 (1995) 563 
\bibitem{Eck95} G. Ecker, Prog. Part. Nucl. Phys. 35 (1995) 1
\bibitem{Be95} V. Bernard, N. Kaiser and U.G. Meissner, Int. J.
Mod. Phys. E4 (1995) 193 
\bibitem{OOP98} J.A. Oller, E. Oset and J.R. Pel\'aez, Phys. Rev.
Lett. 80 (1998) 2452
\bibitem{To71} D.N. Tovee et al., Nucl. Phys. B33 (1971) 493 
\bibitem{No78} R.J. Nowak et al., Nucl. Phys. B139 (1978) 61
\bibitem{Adm81} A.D. Martin, Nucl. Phys. B179 (1981) 33 
\bibitem{Iw97} M. Iwasaki et al., Phys. Rev. Lett. 78 (1997)
3067
\bibitem{nacher98} J.A. Nacher, E. Oset, H. Toki and A. Ramos,
submitted to Phys. Lett. B
\bibitem{moto90} H. Band\={o}, T. Motoba and J. \v{Z}ofka, Int. J.
Mod.  Phys. A5 (1990) 4021 
\bibitem{Nagae98} T. Nagae et al., Phys. Rev. Lett. 80 (1998) 1605
\bibitem{Batty78} C.J. Batty et al, Phys. Lett. 74B (1978) 27,
{\it ibid.}  87B (1979) 324
\bibitem{oset90} E. Oset, P. Fern\'andez de C\'ordoba, 
L.L. Salcedo and R. Brockmann, Phys. Reports 188 (1990) 79 
\bibitem{meirav89} O. Meirav, E. Friedman, R.R. Johnson, 
R. Olszewski and P. Weber, Phys. Rev. C40 (1989) 843 
\bibitem{ramos94} A. Ramos, E. Oset and L.L. Salcedo, Phys. Rev.
C50 (1994) 2314
\bibitem{batty81} C.J. Batty, Nucl. Phys. A372 (1981) 418 
\bibitem{oku99} S. Hirenzaki, Y. Okumura, H. Toki, E. Oset and A.
Ramos, in preparation.
\end{references}
\end{document}